\documentclass[preprintnumbers,amsmath,amssymb,floatfix,12pt,onecolumn,prd,superscriptaddress,nofootinbib,showpacs,showkeys]{revtex4}
\usepackage{graphicx}
\usepackage{epsfig}
\usepackage{bm}
\usepackage{amsfonts}

\def\ben{\begin{equation}}
\def\een{\end{equation}}

 \def\bd{\begin{document}} \def\ed{\end{document}}
\def\ds{\documentstyle} \let\fr=\frac \let\bl=\bigl \let\br=\bigr
\let\Br=\Bigr \let\Bl=\Bigl
\let\bm=\bibitem
\let\na=\nabla
\let\pa=\partial \let\ov=\overline
\newcommand{\be}{\begin{equation}}
\newcommand{\ee}{\end{equation}}
\def\ba{\begin{array}}
\def\ea{\end{array}}
\def\ft#1#2{{\textstyle{\frac{\scriptstyle #1}{\scriptstyle #2} } }}
\def\fft#1#2{{\frac{#1}{#2}}}
\def\del{\partial}
\def\vp{\varphi}
\def\sst#1{{\scriptscriptstyle #1}}
\def\oneone{\rlap 1\mkern4mu{\rm l}}
\def\td{\tilde}
\def\wtd{\widetilde}
\def\ie{{\it i.e.\ }}
\def\dalemb#1#2{{\vbox{\hrule height .#2pt
        \hbox{\vrule width.#2pt height#1pt \kern#1pt
                \vrule width.#2pt}
        \hrule height.#2pt}}}
\def\square{\mathord{\dalemb{6.8}{7}\hbox{\hskip1pt}}}
\newcommand{\ho}[1]{$\, ^{#1}$}
\newcommand{\hoch}[1]{$\, ^{#1}$}
\newcommand{\bea}{\setlength\arraycolsep{2pt} \begin{eqnarray}}
\newcommand{\eea}{\end{eqnarray}}
\newcommand{\ra}{\rightarrow}
\newcommand{\lra}{\longrightarrow}
\newcommand{\Lra}{\Leftrightarrow}
\newcommand{\bp}{\tilde \beta^\prime}
\newcommand{\tr}{{\rm tr} }
\newcommand{\Tr}{{\rm Tr} }
\def\0{{\sst{(0)}}}
\def\1{{\sst{(1)}}}
\def\2{{\sst{(2)}}}
\def\3{{\sst{(3)}}}
\def\4{{\sst{(4)}}}
\def\5{{\sst{(5)}}}
\def\6{{\sst{(6)}}}
\def\7{{\sst{(7)}}}
\def\8{{\sst{(8)}}}
\def\m{{\sst{(m)}}}
\def\n{{\sst{(n)}}}
\def\cA{{{\cal A}}}
\def\cB{{{\cal B}}}
\def\cF{{{\cal F}}}
\def\cG{{{\cal G}}}
\def\cH{{{\cal H}}}
\def\tV{\widetilde V}
\def\tW{\widetilde W}
\def\tH{\widetilde H}
\def\tE{\widetilde E}
\def\tF{\widetilde F}
\def\tA{\widetilde A}
\def\im{{{\rm i}}}
\def\tY{{{\wtd Y}}}
\def\ep{{\epsilon}}
\def\vep{{\varepsilon}}
\def\bD{{{\bar D}}}
\def\R{{{\mathbb R}}}
\def\C{{{\mathbb C}}}
\def\H{{{\mathbb H}}}
\def\CP{{{\mathbb C}{\mathbb P}}}
\def\RP{{{\mathbb R}{\mathbb P}}}
\def\Z{{{\mathbb Z}}}
\def\bA{{{\mathbb A}}}
\def\bB{{{\mathbb B}}}
\def\bC{{{\mathbb C}}}
\def\bD{{{\mathbb D}}}
\def\bE{{{\mathbb E}}}
\def\bZ{{{\mathbb Z}}}
\def\Re{{{\frak{Re}}}}
\def\Im{{{\frak{Im}}}}
\def\cosec{{\,\hbox{cosec}\,}}
\def\Gm{{\Gamma_{\!\! -}}}
\def\Gp{{\Gamma_{\!\! +}}}
\def\stan{{standard }}
\def\nonstan{{supernumerary }}
\def\p{{\partial}}
\def\kdel#1{{\fft{\del}{\del#1}}}

\def\bog{{Bogomolny }}
\def\om{{\omega}}

\newcommand{\nnr}{\nonumber \\}
\newcommand{\pd}{\partial}
\newcommand{\ud}{\textrm{d}}
\newcommand{\dTH}{T^{\prime \, 0}_\textrm{H}}
\newcommand{\dOi}{\Omega^{\prime \, 0}_i}
\newcommand{\bx}{{\bf x}}
\begin{document}

\title{A note on  holographic superconductors with Weyl Corrections}

\author{\textbf{D. Momeni}}
\email{d.momeni@yahoo.com}
 \affiliation{Department of Physics , Faculty of Sciences,
  Tarbiat Moallem University, Tehran , Iran}

\author{\textbf{M.R. Setare}}
\email{Rezakord@ipm.ir}
 \affiliation{Department of Campus of Bijar, University of  Kurdistan , Bijar, IRAN}

\begin{abstract}
We study analytical properties of the holographic  superconductors
with Weyl corrections. We describe the phenomena in the probe limit
neglecting backreaction of the space-time. We observe that for the
conformal dimension  $\triangle_{+}=3$, the minimum value of the
critical temperature $T^{Min}_c$ at which condensation sets, can be
obtained directly from the equations of motion as $T^{Min}_c\approx
0.170845\sqrt[3]{\rho}$, which is in very good agreement with the
numerical value $T^{Min}_c=0.170\sqrt[3]{\rho}$
[Phys.Lett.B697:153-158,2011]. This value of $T^{Min}_c$ corresponds
to the value of the Weyl's coupling $\gamma=-0.06$ in table (1) of
[Phys.Lett.B697:153-158,2011]. We calculate the $T^{Min}_c\approx
0.21408\sqrt[3]{\rho}$ for another Weyl's coupling $\gamma=0.02$ and
the the conformal dimension $\triangle_{-}=1$. Further, we show that
the critical exponent is $\beta=\frac{1}{2}$. We observe that there
is a linear relation between the charge density $\rho$ and the
chemical potential difference $\mu-\mu_c$ qualitatively matches the
numerical curves.

\end{abstract}
\pacs{04.70.Bw, 11.25.Tq, 74.20.-z}
 \keywords{Classical Black holes; Gauge/string duality; High-$T_C$ superconductors theory}
 \newpage
 \maketitle
\section{Introduction}
The anti de Sitter/conformal field theory (AdS/CFT) correspondence
\cite{maldacena} provides a powerful theoretical method to
investigate the strongly coupled field theories. It may have useful
applications in condensed matter physics, especially for studying
scale-invariant strongly-coupled systems, for example, low
temperature systems near quantum criticality (see for example
\cite{condencesd1,condencesd2} and references therein). Recently, it
has been proposed that the AdS/CFT correspondence also can be used
to describe superconductor phase transition \cite{super1,super2}.
Since the high $T_c$ superconductors are shown to be in the strong
coupling regime,the BCS theory fails and one expects that the
holographic method could give some insights into the pairing
mechanism in the high $T_c$ superconductors. Various holographic
superconductors have been studied in Einstein theory \cite{GR1,GR2}
or extended versions as Gauss-Bonnet (GB)\cite{GB1,GB2,GB3,GB4} and
even in Horava-Lifshitz theory \cite{HL1,HL2}. AdS/CFT can also
describe superfluid states in which the condensing operator is a
vector and hence rotational symmetry is broken, that is, p-wave
superfluid states \cite{pwave1,pwave2,pwave3}. Here the CFT has a
global $SU(2)$ symmetry and hence three conserved currents
$J^{\mu}_a$ , where $a = 1, 2, 3$ label the generators of $SU(2)$.
All these works are based on a numerical analysis of the equations
on motion (EOM) near the horizon and the asymptotic limit by a
suitable shooting method. But as we know the analytical methods are
better and easy for invoking in different problems. The first
pioneering work on analytic methods in this topic was the Hertzog's
work \cite{herzog}. He showed that at least in probe limit, by
solving equations analytically, one can obtain the critical exponent
and the expectation values of the dual operators. The philosophy
hidden behind his calculations is perturbation theory. Near the
critical point the value of the scalar field $\psi$ is small and
consequently we can treat the expectation values of the dual
boundary operators $\epsilon\equiv<O_{\Delta_{\pm}}>$ as a
perturbation parameter. This method has been used recently by Kanno
for investigating the GB superconductors even away from the probe
limit \cite{kanno}. Applying the analytical methods has been
extended to new trends (see for example \cite{analytic} and the
references in it).  In \cite{analytic} the authors have shown that
one can obtain the critical exponent and the critical temperature by
applying a variational method on the EOM. Their method and
terminology is simple and very sound. Instead of involving in
numerical problems , we can obtain the critical temperature $T_c$
and the exponent of the criticality very easily
 by computing a simple variational approach. They studied different
  modes of super criticality
s-wave, p-wave and even d-wave. Thus as we know, there is two
major method for analytical study of superconductors:\emph{ The
small parameter perturbation theory} as used by \cite{herzog} and
the \emph{variational method} \cite{analytic}. We must attend
that, the variational method, which has been used in the present
work, gives only the minimum value of the critical temperature
$T^{Min}_c$ for a model with a typical parameter. For example if
we focus on Weyl corrections to holographic superconductors, as
it has been shown in \cite{weyl}, for a large range of the
coupling value $-\frac{1}{16}<\gamma<\frac{1}{24}$, there is a
universal relation for the critical temperature $T_c\simeq
\sqrt[3]{\rho}$. The proportionality constant depends on the Weyl
coupling $\gamma$ and can be computed using the numerical
methods. There is a low limit of superconductivity with critical
temperature $T^{Min}_c=0.170\sqrt[3]{\rho}$ corresponds to the
value $\gamma=-0.06$.
 Specially recently there are many interests
on GB and Weyl corrected superconductors which in them, one is
working with a corrected BH. The holographic superconductors with
Weyl corrections has been studied recently \cite{weyl}. They studied
the problem numerically. Our program in this paper is studying the
Weyl corrections to the superconductors analytically. Our plan is
organized as follows. In section 2, we construct the basic model of
the 3+1 holographic superconductor with Weyl corrections. In section
3 we present the analytical results for the condensation and minimum
value of the critical temperature for different scaling and  the
critical exponent $\beta$ via variational bound. Conclusions and
discussions follow in section 4.
 \\

\section{Weyl corrected s-wave superconductors}
The s-wave holographic superconductors can be constructed from a
U(1) scalar gauge field coupled to a massive charged   scalar
field (complex field ). The simplest form of the action in five
dimensions (3+1 holographic picture)  with Weyl corrections, in
units in which the AdS radius L=1, charge $e=1$,
reads\cite{weylcorrection}

\begin{eqnarray}
 S=\int
dtd^{4}x\sqrt{-g}\{\frac{1}{16\pi
G_{5}}(R+12)-\frac{1}{4}(F^{\mu\nu}F_{\mu\nu}-4\gamma
C^{\mu\nu\rho\sigma}F_{\mu\nu}F_{\rho\sigma})-|D_{\mu}\psi|^2-m^2|\psi|^2\}
\end{eqnarray}
Here $G_{5}$ is the gravitational constant, the (12) term gives
the negative cosmological constant,
$F_{\mu\nu}=\partial_{\mu}A_{\nu}-\partial_{\nu}A_{\mu}$ and as
the usual $D_{\mu}=\nabla_{\mu}-iA_{\mu}$. The gauge field
$A_{\mu}$ lives in bulk and produces a conserved current
$J^{\mu}$. This current corresponds to a \emph{global U(1)
symmetry}.

 About the action (1) we can
say that  \emph{since the background geometry will be an Einstein
metric, we will argue that there is a unique tensorial structure
correcting the Maxwell term at leading order in derivatives, arising
from a coupling to the Weyl tensor and leading to the dimension-six
operator in (1) parametrized by the constant $\gamma$. Other
curvature couplings simply provide constant shifts when considering
linearized gauge field fluctuations about the background}. There is
another reason for considering the Weyl correction, which is related
to the quantum corrections: In any background in which additional
charged matter fields are integrated out below their mass threshold,
the Weyl coupling $C^{\mu\nu\rho\sigma}F_{\mu\nu}F_{\rho\sigma}$ is
generated at 1-loop, with a coefficient $\gamma=\frac{\alpha}{m^2}$
first computed (for four dimension) by Drummond and Hathrell
\cite{qc}.

The Weyl's coupling $\gamma$ is limited such that it's value is in
the interval $-\frac{1}{16}<\gamma<\frac{1}{24}$. In probe limit ,
we neglect from the back reactions and in this case, the gravity
sector is effectively decoupled from the matter field's sector. In
this probe limit, the exact solution for Einstein-Yang Mills
equations is b black brain given by
\begin{eqnarray}
 ds^2=r^2(-fdt^2+dx^idx_i)+\frac{dr^2}{r^2f}
\end{eqnarray}
Here
\begin{eqnarray}
f=1-(\frac{h}{r})^4
\end{eqnarray}
and the a horizon locates at $r=h$. This solution is asymptotically
anti-de Sitter. The temperature of the dual conformal field theory
(CFT) is nothing just the Hawking temperature  and reads
$T=\frac{h}{\pi}$. When the temperature of the black brane falls
below a critical value $T_c$ , it happens a phase transition between
the normal phase and a new phase , in which the scalar field $\psi$
condenses. If the model has such a solution, we state that our field
theory has a superfluid phase.

We choose a gauge as $\psi=\psi(r),A_{t}=\varphi(r)$. It is more
conveint to work in terms of the dimensionless parameter
$\xi=\frac{h}{r}$, in which the horizon is $\xi=1$ and the boundary
at infinity locates at $\xi=0$. The resulting Yang-Mills equations
\begin{eqnarray}\nonumber
D_{\mu}D^{\mu}\psi-m^2\psi=0 \\\nonumber
\frac{1}{\sqrt{-g}}\partial_{\mu}(\sqrt{-g}F^{\mu\nu})=i[\psi^{*}D^{\nu}\psi-\psi
D^{\nu *}\psi^{*}]
\end{eqnarray}

for metric (2) are given by
\begin{eqnarray}
 \psi''-\frac{\xi^4+3}{\xi(1-\xi^4)}\psi'+(\frac{\varphi^2}{h^2(1-\xi^4)^{2}}-\frac{m^2}{\xi^2(1-\xi^4)})\psi=0\\
 (1-24\gamma \xi^4)\varphi''-(\frac{1}{\xi}+72\gamma
 \xi^3)\varphi'-\frac{2\psi^2}{\xi^2(1-\xi^4)}\varphi=0
 \end{eqnarray}
 where prime now denotes derivative with respect to $\xi$. Also we
 fix the mass of the scalar field to $m^2=-3$ which is obviously
 above the  Breitenlohner-Freedman bound \cite{BF}.
The adequate and sufficient boundary conditions for these equations
can be written on horizon $\xi=1$, the bulk's boundary $\xi=0$. On
the horizon we've $\varphi(1)=0,\psi'(1)=\frac{2}{3}\psi(1)$ and on
the boundary of bulk, the following asymptotic forms of the
solutions must be existed
\begin{eqnarray}
 \varphi\approx \mu-\frac{\rho}{h} \xi^2\\
 \psi
 \approx\frac{<O_{\Delta_{\pm}}>}{\sqrt{2}h^{\Delta_{\pm}}}\xi^{\Delta_{\pm}}
 =\psi^{(1)}\xi^{\Delta_{+}}+\psi^{(3)}\xi^{\Delta_{-}}
 \end{eqnarray}
$\mu$ and $\rho$ are dual to the chemical potential and charge
density of the boundary CFT, $\psi^{(1)}$ and $\psi^{(3)}$ are dual
to the source and expectation value of the boundary operator $O $
respectively and $<O_{\Delta_{\pm}}>$ are the condensation with
dimension $\Delta_{\pm}$ where the dimension $\Delta_{\pm}$ is given
by
\begin{eqnarray}
\Delta_{\pm}=\{3,1\}
\end{eqnarray}
The conformal scaling dimension $\Delta\geq1$ is related to the mass
and the de Sitter radius by $m^2L^2=\triangle(\triangle-4)$.

\section{Analytical results for the condensation and critical temperature}
We know that there is a second order continuous phase transition at
the critical temperature, the solution of the EOMs (4,5) at  $T_c$
is
 \begin{eqnarray}
\varphi=\lambda h_c(1-\xi^2)
\end{eqnarray}
where $h_c$ is the radius of the horizon at $T= T_c$.
 As $T\rightarrow T_c$, the scalar filed's EOM tends to the
 following form
\begin{eqnarray}
 -\psi''+\frac{\xi^4+3}{\xi(1-\xi^4)}\psi'-\frac{3}{\xi^2(1-\xi^4)}\psi=\frac{\lambda^2}{(1+\xi^2)^2}\psi,
\end{eqnarray}

here $\lambda=\frac{\rho}{h_c^{3}}$. By solving the equation of
(10), we can obtain the value of $T_c$. To match the behavior at
the boundary, we can define
\begin{eqnarray}
  \psi(\xi)=
  \frac{<O_{\Delta_{\pm}}>}{\sqrt{2}h^{\Delta_{\pm}}}\xi^{\Delta_{\pm}}\Omega(\xi)
 \end{eqnarray}

where, according to eq.(7), $\Omega$ is normalized as $\Omega (0) =
1$. We deduce

\begin{eqnarray}
-\Omega''+\frac{\Omega'}{\xi}(\frac{\xi^4+3}{1-\xi^4}-2\Delta_{\pm})+\frac{\Delta_{\pm}
^2
\xi^4-(\Delta_{\pm}-1)(\Delta_{\pm}-3)}{\xi^2(1-\xi^4)}\Omega=\frac{\lambda^2}{(1+\xi^2)^2}\Omega
 \end{eqnarray}
 when $z\rightarrow 0$,
$\frac{\Omega'}{\xi}$ should be finite, so this equation is to be
solved subject to the boundary condition $\Omega'(0)=0$.

\subsection{Variational approach}

 Now we use
from the variation method to solve the Sturm-Liouville problem (12).
The Sturm-Liouville eigenvalue problem is to solve the equation
\cite{analytic}
\begin{eqnarray}
\frac{d }{d\xi} [k(\xi) \frac{d\Omega}{d\xi}  ] - q(\xi)\Omega(\xi)
+ \lambda^2\rho(\xi)\Omega(\xi) = 0
 \end{eqnarray}
 with boundary condition
\begin{eqnarray}
k(\xi)\Omega(\xi)\Omega'(\xi)|^1 _0 = 0
\end{eqnarray}
 The Sturm-Liouville
problem can be result to be a functional minimize problem
\begin{eqnarray}
F [\Omega(\xi)] = \frac{\int ^1 _0 d\xi(k(\xi)\Omega ' (\xi) ^2 +
q(\xi)\Omega(\xi) ^2)}{\int ^1 _0 d\xi\rho(\xi)\Omega(\xi)^ 2 }
 \end{eqnarray}
Then  th eigenvalue $\lambda_n$ can also be obtained by variation
Eq. (14). This eigenvalue is the minimum value of a sequence of
the eigenvalues $\{\lambda_n\}^{\infty}_{0}$ i.e. we obtain
$\lambda_{0}<\lambda_n$. It is a familiar result from the
functional theory. For Eq.(12) we immediately obtain
\begin{eqnarray}
k(\xi)=\xi^{2\Delta_{\pm}-3}(1-\xi^4)
\\
q(\xi)=-\xi^{2\Delta_{\pm}-5}(\Delta_{\pm} ^2
\xi^4-(\Delta_{\pm}-1)(\Delta_{\pm}-3))\\
\rho(\xi)=\frac{\xi^{2\Delta_{\pm}-3}(1-\xi^2)}{1+\xi^2}
 \end{eqnarray}
The boundary condition (14) is very serious for our computational
settings. We will discuss two different cases $\Delta_{\pm}=\{3,1\}$
separately. In each case, we will chose different kind of the trial
function $\Omega(\xi)$.

\subsubsection{\textbf{Case} $\Delta_{+}=3$}
If we fix $\Delta_{+}=3$, then the boundary condition (14) reads as
\begin{eqnarray}
\xi^3(1-\xi^4)\Omega(\xi)\Omega'(\xi)|^1 _0 = 0
\end{eqnarray}
In right boundary point $\xi=1$ it is satisfied clearly. For left
point $\xi=0$ we must be careful. Indeed to have
  $\lim_{\xi\rightarrow0}\xi^3(1-\xi^4)\Omega(\xi)\Omega'(\xi)=0$, the
eigenfunction $\Omega(\xi)$ must be non singular. Additionally, we
impose the auxiliary conditions on it's value at left point of the
interval $[0,1]$. In order to use the variation method , we have to
specifies the trial eigenfunction $\Omega(\xi)$. We impose the
auxiliary boundary conditions $\Omega(0) = 1$ and $\Omega'(0)=0$.
The third order trial eigenfunction is then
\begin{figure}
\centering
 \includegraphics[width=10cm,angle=0] {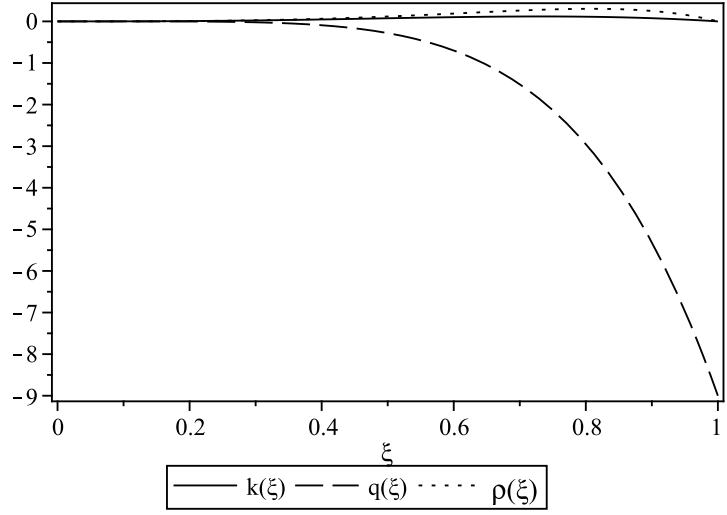}% scale goes from 0 to 1.
  \caption{Variation of the functions $k(\xi),q(\xi),\rho(\xi)$ for the case $\Delta_{+}=3$}
 \label{1.eps}
\end{figure}

\begin{eqnarray}
\Omega(\xi)=1-\alpha\xi^2+\beta\xi^3
 \end{eqnarray}
We obtain
\begin{eqnarray}
|\lambda_{\alpha,\beta}|^2=\frac{-0.633333 \alpha ^2+1.01299 \alpha
\beta
   +2.25 \alpha -0.375 \beta ^2-2 \beta
   -1.5}{0.0151862 \alpha ^2+\alpha  (-0.0241323
   \beta -0.052961)+0.00981385 \beta ^2+0.0393597
   \beta +0.0568528}
 \end{eqnarray}
which attains its minimum at $\alpha=-3.92548,\beta=4.76575$. We
obtain
\begin{eqnarray}
|\lambda_{-3.92548,4.76575}|^2\approx 41.9572
\end{eqnarray}
which can be compared with the numerical value \cite{weyl}.  The
figure (1) shows the the functions $k(\xi),q(\xi),\rho(\xi)$ for the
case $\Delta_{+}=3$.

The  \emph{Minimum} critical temperature $T^{Min}_c$ is
\begin{eqnarray}
T^{Min}_c=\frac{h_c}{\pi}=\frac{1}{\pi}\sqrt[3]{\frac{\rho}{\lambda}}
\end{eqnarray}
so for $\Delta_{+} = 3$, $T^{Min}_c\approx 0.170845\sqrt[3]{\rho}$,
which is in very good agreement with the numerical value
$T^{Min}_c=0.170\sqrt[3]{\rho}$ for $\gamma=-0.06$ of table (1) in
\cite{weyl}. In fact, this analytical calculation can be done even
better if we include higher order of $\xi$. However, for qualitative
analyze, the third order trial eigenfunction is good enough and we
use it.

\subsubsection{\textbf{Case} $\Delta_{-}=1$}
Now we fix $\Delta_{-}=1$, then the boundary condition (14) reads as
\begin{eqnarray}
\frac{(1-\xi^4)}{\xi}\Omega(\xi)\Omega'(\xi)|^1 _0 = 0
\end{eqnarray}

The second order trial eigenfunction satisfying the $\Omega(0) =0$
is then

\begin{eqnarray}
\Omega(\xi)=a\xi+b\xi^2+c\xi^3
 \end{eqnarray}
Now we obtain

\begin{eqnarray}
\lim_{\xi\rightarrow0}\frac{1-\xi^4}{\xi}(a\xi+b\xi^2+c\xi^3)(a+2b\xi+3c\xi^2)=0
 \end{eqnarray}
It results $a=0$. Now we examine $\Omega(\xi)=b\xi^2+c\xi^3$ in
functional (15) which attains its minimum at $b=0.465,c=-0.323$. We
obtain

\begin{eqnarray}
|\lambda_{0.465,-0.323}|^2\approx 10.806
\end{eqnarray}

So for $\Delta_{-} = 1$, $T^{Min}_c\approx 0.21408\sqrt[3]{\rho}$.
This value for critical temperature corresponds to the value of the
$\gamma=0.02$ from numerical solving given by \cite{weyl}. The
figure (2) shows the variation of $k(\xi),q(\xi),\rho(\xi)$ with
respect to $\xi$.

\begin{figure}
\centering
 \includegraphics[width=10cm,angle=0] {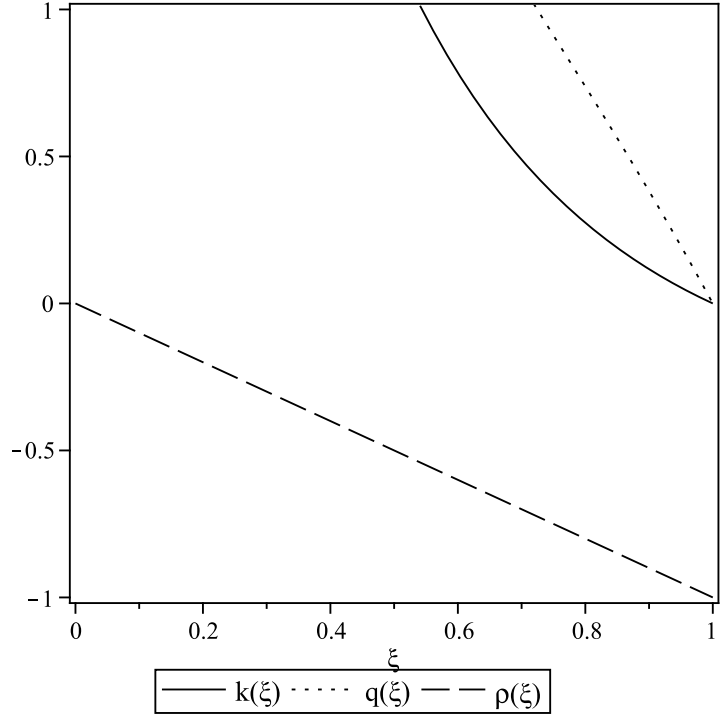}% scale goes from 0 to 1.
  \caption{Variation of the functions $k(\xi),q(\xi),\rho(\xi)$ for the case $\Delta_{-}=1$}
 \label{2.eps}
\end{figure}

\subsection{Critical exponent $\beta$} Now we begin to solve the
equation for $\varphi$ to obtain the behavior of the order parameter
at $T_c$. Away from (but close to) the critical temperature, the
field eq.(5) for $\varphi$ becomes

\begin{eqnarray}
 (1-24\gamma \xi^4)\varphi''-(\frac{1}{\xi}+72\gamma
 \xi^3)\varphi'\approx[\frac{<O_{\Delta_{\pm}}>^2}{h^{2\Delta_{\pm}}}
 \frac{\xi^{2\Delta_{\pm}-2}}{1-\xi^4}\Omega(\xi)^2]\varphi
 \end{eqnarray}

where the parameter
 $\varepsilon^2=\frac{<O_{\Delta_{\pm}}>^2}{h^{2\Delta_{\pm}}}$ is
small.  Because near the critical chemical $\mu_c$ , the
condensation of the operator is very small, we can expand $\varphi$
in $\varepsilon$ as \cite{cai}
\begin{eqnarray}
\varphi(\xi)\sim \mu_c+\varepsilon \chi(\xi)
\end{eqnarray}
where $\chi(\xi)$ is the general correction function to be
$\chi(0)=1$. The equation of motion of $\chi(\xi)$ is
\begin{eqnarray}
\chi''(\xi)-\frac{\frac{1}{\xi}+72\gamma
 \xi^3}{1-24\gamma \xi^4}\chi'(\xi)=\varepsilon\mu_c\frac{\xi^{2\Delta_{\pm}-2}}{(1-\xi^4)(1-24\gamma \xi^4)
 }\Omega(\xi)^2
\end{eqnarray}
Multiplying
\begin{eqnarray}\nonumber
\eta(\xi)={\frac {\sqrt [4]{-1+24\,\gamma\,{\xi}^{4}}{{\rm
e}^{-3\,\sqrt {6\gamma }{\it arctanh} \left( 2\,\sqrt
{6\gamma}{\xi}^{2}
 \right) }}}{\xi}}
\end{eqnarray}
  to both sides of the above equation the equation of $\chi(\xi)$
is reduced to
\begin{eqnarray}
\frac{d }{d\xi} [\eta(\xi) \frac{d\chi}{d\xi}  ]=
-\varepsilon\mu_c\frac{\xi^{2\Delta_{\pm}-3}{{\rm e}^{-3\,\sqrt
{6\gamma}{\it arctanh} \left( 2\,\sqrt { 6\gamma}{\xi}^{2} \right)
}}\Omega(\xi)^2 }{(-1+24\gamma \xi^4)^{3/4}(1-\xi^4)}
 \end{eqnarray}
Making integration of both sides, we get
\begin{eqnarray}
\eta(\xi) \frac{d\chi}{d\xi}|_{0}^{1}
=-\varepsilon\mu_c\int_{0}^{1}\frac{\xi^{3}{{\rm e}^{-3\,\sqrt
{6\gamma}{\it arctanh} \left( 2\,\sqrt { 6\gamma}{\xi}^{2} \right)
}}(1-\alpha\xi^2)^2 }{(-1+24\gamma \xi^4)^{3/4}(1-\xi^4)}d\xi
 \end{eqnarray}
where we have used the trial function $\Omega(\xi)=1-\alpha\xi^2$,we
fixed $\Delta_{+}=3$. Near $\xi=0$, $\varphi(\xi)$ can be expanded
as
\begin{eqnarray}
 \varphi\approx \mu-\frac{\rho}{h}
 \xi^2\approx\mu_c+\varepsilon(\chi(0)+\chi'(0)\xi+\frac{1}{2}\chi''(0)\xi^2+...)
 \end{eqnarray}
Comparing the coefficients of $\xi^0$ term in both sides of the
above formula, we get
\begin{eqnarray}
\mu-\mu_c\approx\varepsilon\chi(0)
\end{eqnarray}
Besides, from the $\xi^1$ term in (31), we obtain that $\chi'(0)=0$.
Therefore, from the equation (29) and the boundary conditions of
$\chi(z)$, we can solve $\chi(\xi)$ to be
\begin{equation}
\chi(\xi)=-\varepsilon
\mu_c\{c'+\frac{c\xi^2}{2}+9c\gamma\xi^4+\{\frac{-1}{24}+c\gamma(1+108\gamma)\}\xi^6+O(\xi^7)\}
\end{equation}
 Here, $c,c'$ both are the integration constants. Thus we obtain $\chi(0)\approx-\varepsilon  \mu_c c'$. Further we
have
\begin{eqnarray}
\mu-\mu_c\approx-  \mu_c
c'\varepsilon^2\\<O_{\Delta_{\pm}}>\approx
\frac{h^3}{\sqrt{-c'\mu_c}}\sqrt{\mu-\mu_c}
\end{eqnarray}
This critical exponent $\frac{1}{2}$ for the condensation value and
$\mu-\mu_c$ qualitatively match the numerical curves in Figure.1 of
Ref.\cite{weyl}. Further we can show that there is a linear relation
between the charge density $\rho$ and the chemical potential
difference $\mu-\mu_c$ qualitatively matches the numerical curves in
\cite{weyl}. Moreover, this linear relation between $\rho$ and
$\mu-\mu_c$ can also be frequently seen in the numerical analysis .

\section{Conclusions}
In this paper, we have studied the analytical properties of the
s-wave  holographic superconductor phase transitions with the Weyl
corrections and obtained the analytical solutions of this model for
the scalar operators of conformal dimension $\Delta=\{3,1\}$.
Actually, we have analytically obtained the minimum bound for
critical temperature $T^{Min}_c$ in s-wave model. We found that the
critical temperature $T_c\approx 0.170845\sqrt[3]{\rho}$ for
$\Delta=3$ and $T^{Min}_c\approx 0.21408\sqrt[3]{\rho}$ for
conformal dimension $\Delta=1$ which both are perfectly in agreement
with the previous numerical values \cite{weyl}. We found that the
critical exponent of condensation operator is always
$\beta=\frac{1}{2}$ in this model. Also , we  obtained the linear
relations between the charge density $\rho$ and the chemical
potential difference $\mu-\mu_c$, which is also qualitatively
consistent with the previous numerical results.
\section{Acknowledgment}
The authors would like to thank Jian Pin Wu (Beijing Normal
University-China) for his helpful discussions and comments.

\end{document}